\input harvmac

\input epsf
%
\noblackbox

\newcount\figno
\figno=0
\def\fig#1#2#3{
\par\begingroup\parindent=0pt\leftskip=1cm\rightskip=1cm\parindent=0pt
\baselineskip=11pt
\global\advance\figno by 1
\midinsert
\epsfxsize=#3
\centerline{\epsfbox{#2}}
\vskip 12pt
{\bf Fig.\ \the\figno: } #1\par
\endinsert\endgroup\par
}
\def\figlabel#1{\xdef#1{\the\figno}}
\def\encadremath#1{\vbox{\hrule\hbox{\vrule\kern8pt\vbox{\kern8pt
\hbox{$\displaystyle #1$}\kern8pt}
\kern8pt\vrule}\hrule}}

\def\frac#1#2{{#1 \over #2}}

\def\p{\partial}
\def\semi{\subset\kern-1em\times\;}
\def\bar#1{\overline{#1}}

\def\p{\partial}

\def\h{{1 \over 2}}

\font\zfont = cmss10 
\font\zfonteight = cmss8 
\def\ZZ{\hbox{\zfont Z\kern-.4emZ}}
\def\ZZs{\hbox{\zfonteight Z\kern-.4emZ}}
%

\Title{\vbox{\baselineskip12pt
\hbox{hep-th/0105227}
\hbox{EFI-01-15}
\hbox{UCLA/01/TEP/9}
\vskip-.5in}}
{\vbox{\centerline{Loop Corrected Tachyon Condensation}}}
\medskip\bigskip


\centerline{\it{ 
Ben Craps${}^{1}$, Per Kraus${}^{2}$ and 
Finn Larsen${}^{1}$} \footnote{$ ^\star$}{Address after July 1, 2001:
Michigan Center for Theoretical Physics, 
University of Michigan, Ann Arbor, MI-48109.}}

\bigskip
\centerline{${}^1$Enrico Fermi Institute, University of Chicago,
5640 S. Ellis Av., Chicago, IL-60637}
\centerline{${}^2$Department of Physics, University of California, 
Los Angeles, CA 90095}

\medskip
\baselineskip18pt
\medskip\bigskip\medskip\bigskip\medskip
\baselineskip16pt
\noindent
We study loop corrections in boundary string field theory (BSFT). 
After commenting on problems with quantizing the tree level BSFT as an 
ordinary field theory, we discuss the tree level coupling to closed 
strings and define the loop corrections via factorization in the closed 
string channel. This description is weakly coupled in the vicinity of 
the closed string vacuum. Our proposal for the one-loop effective action 
differs in general from  computing the  annulus or
 cylinder partition functions. 
We also compute the decay rates and the loop corrections to the tensions of 
unstable branes in perturbative string theory.

\Date{May, 2001}

\lref\StromingerZD{
A.~Strominger,
``Closed Strings In Open String Field Theory,''
Phys.\ Rev.\ Lett.\  {\bf 58}, 629 (1987).
}

\lref\SrednickiRV{
M.~Srednicki and R.~P.~Woodard,
``Closed From Open Strings In Witten's Theory,''
Nucl.\ Phys.\ B {\bf 293}, 612 (1987).
}

\lref\HarveyJT{
J.~A.~Harvey, P.~Kraus, F.~Larsen and E.~J.~Martinec,
``D-branes and strings as non-commutative solitons,''
JHEP {\bf 0007}, 042 (2000)
[hep-th/0005031].
}

\lref\GibbonsHF{
G.~Gibbons, K.~Hori and P.~Yi,
``String fluid from unstable D-branes,''
Nucl.\ Phys.\ B {\bf 596}, 136 (2001)
[hep-th/0009061].
}

\lref\SenKD{
A.~Sen,
``Fundamental strings in open string theory at the tachyonic vacuum,''
hep-th/0010240.
}

\lref\SenXM{
A.~Sen,
``Universality of the tachyon potential,''
JHEP {\bf 9912}, 027 (1999)
[hep-th/9911116].
}

\lref\WeinbergVP{
E.~J.~Weinberg and A.~Wu,
``Understanding Complex Perturbative Effective Potentials'',
Phys.\ Rev.\ D {\bf 36}, 2474 (1987).
}

\lref\GuthYA{
A.~H.~Guth and S.~Pi,
``The Quantum Mechanics Of The Scalar Field In The New Inflationary 
Universe'',
Phys.\ Rev.\ D {\bf 32}, 1899 (1985).
}

\lref\MarcusVS{
N.~Marcus,
``Unitarity And Regularized Divergences In String Amplitudes'',
Phys.\ Lett.\ B {\bf 219}, 265 (1989).
}

\lref\BanksQS{
T.~Banks and E.~Martinec,
``The Renormalization Group And String Field Theory'',
Nucl.\ Phys.\ B {\bf 294}, 733 (1987).
}

\lref\LiuNZ{
J.~Liu and J.~Polchinski,
``Renormalization Of The Mobius Volume'',
Phys.\ Lett.\ B {\bf 203}, 39 (1988).
}

\lref\Freedman{
D.~Z.~Freedman, S.~B.~Giddings, J.~A.~Shapiro and C.~B.~Thorn,
``The Nonplanar One Loop Amplitude In Witten's String Field Theory'',
Nucl.\ Phys.\ B {\bf 298}, 253 (1988).
}

\lref\cmnp{
A.~Cohen, G.~Moore, P.~Nelson and J.~Polchinski,
``An Off-Shell Propagator For String Theory'',
Nucl.\ Phys.\ B {\bf 267}, 143 (1986).}

\lref\AlishahihaTG{
M.~Alishahiha,
``One-loop correction of the tachyon action in boundary superstring field  
theory'',
[hep-th/0104164].
}

\lref\KosteleckyNT{
V.~A.~Kostelecky and S.~Samuel,
``On A Nonperturbative Vacuum For The Open Bosonic String'',
Nucl.\ Phys.\ B {\bf 336}, 263 (1990).
}

\lref\ViswanathanCS{
K.~S.~Viswanathan and Y.~Yang,
``Tachyon condensation and background independent superstring field  
theory'',
[hep-th/0104099].
}

\lref\BardakciCK{
K.~Bardakci and A.~Konechny,
``Tachyon condensation in boundary string field theory at one loop'',
[hep-th/0105098].
}

\lref\CuomoDE{
F.~Cuomo, R.~Marotta, F.~Nicodemi, R.~Pettorino, F.~Pezzella and G.~Sabella,
``Off-shell tachyon amplitudes: Analyticity and projective invariance'',
[hep-th/0011071].
}

\lref\FrolovNB{
S.~A.~Frolov,
``On off-shell structure of open string sigma model'',
[hep-th/0104042].
}

\lref\KrausNJ{
P.~Kraus and F.~Larsen,
``Boundary string field theory of the D D-bar system,''
Phys.\ Rev.\ D {\bf 63}, 106004 (2001)
[hep-th/0012198].
}

\lref\TakayanagiRZ{
T.~Takayanagi, S.~Terashima and T.~Uesugi,
``Brane-antibrane action from boundary string field theory,''
JHEP {\bf 0103}, 019 (2001)
[hep-th/0012210].
}

\lref\NiarchosSI{
V.~Niarchos and N.~Prezas,
``Boundary superstring field theory'',
[hep-th/0103102].
}

\lref\MarinoQC{
M.~Marino,
``On the BV formulation of boundary superstring field theory'',
[hep-th/0103089].
}

\lref\ChalmersDJ{
G.~Chalmers,
``Open string decoupling and tachyon condensation'',
hep-th/0103056.
}

\lref\TakayanagiRZ{
T.~Takayanagi, S.~Terashima and T.~Uesugi,
``Brane-antibrane action from boundary string field theory'',
JHEP {\bf 0103}, 019 (2001)
[hep-th/0012210].
}

\lref\KrausNJ{
P.~Kraus and F.~Larsen,
``Boundary string field theory of the D D-bar system'',
Phys.\ Rev.\ D {\bf 63}, 106004 (2001)
[hep-th/0012198].
}

\lref\SenKD{
A.~Sen,
``Fundamental strings in open string theory at the tachyonic vacuum'',
[hep-th/0010240].
}

\lref\KutasovAQ{
D.~Kutasov, M.~Marino and G.~Moore,
``Remarks on tachyon condensation in superstring field theory'',
[hep-th/0010108].
}

\lref\GerasimovZP{
A.~A.~Gerasimov and S.~L.~Shatashvili,
``On exact tachyon potential in open string field theory'',
JHEP {\bf 0010}, 034 (2000)
[hep-th/0009103].
}

\lref\KutasovQP{
D.~Kutasov, M.~Marino and G.~Moore,
``Some exact results on tachyon condensation in string field theory'',
JHEP {\bf 0010}, 045 (2000)
[hep-th/0009148].
}

\lref\ZwiebachQJ{
B.~Zwiebach,
``Quantum open string theory with manifest closed string factorization'',
Phys.\ Lett.\ B {\bf 256}, 22 (1991).
}

\lref\ZwiebachFE{
B.~Zwiebach,
``Oriented open-closed string theory revisited'',
Annals Phys.\  {\bf 267}, 193 (1998)
[hep-th/9705241].
}

\lref\TseytlinMT{
A.~A.~Tseytlin,
``Sigma model approach to string theory effective actions with tachyons'',
[hep-th/0011033].
}

\lref\TseytlinMV{
A.~A.~Tseytlin,
``Renormalization Group And String Loops'',
Int.\ J.\ Mod.\ Phys.\ A {\bf 5}, 589 (1990).
}

\lref\TseytlinMW{
A.~A.~Tseytlin,
``String Theory Effective Action: String Loop Corrections'',
Int.\ J.\ Mod.\ Phys.\ A {\bf 3}, 365 (1988).
}

\lref\TseytlinRR{
A.~A.~Tseytlin,
``Sigma Model Approach To String Theory'',
Int.\ J.\ Mod.\ Phys.\ A {\bf 4}, 1257 (1989).
}

\lref\FradkinPQ{
E.~S.~Fradkin and A.~A.~Tseytlin,
``Effective Field Theory From Quantized Strings'',
Phys.\ Lett.\ B {\bf 158}, 316 (1985).
} 

\lref\AndreevCB{
O.~D.~Andreev and A.~A.~Tseytlin,
``Partition Function Representation For The Open Superstring 
Effective Action:
Cancellation Of Mobius Infinities And Derivative Corrections To
Born-Infeld Lagrangian'',
Nucl.\ Phys.\ B {\bf 311}, 205 (1988).
}

\lref\ShatashviliPS{
S.~L.~Shatashvili,
``On the problems with background independence in string theory'',
[hep-th/9311177].
}

\lref\ShatashviliKK{
S.~L.~Shatashvili,
``Comment on the background independent open string theory'',
Phys.\ Lett.\ B {\bf 311}, 83 (1993)
[hep-th/9303143].
}

\lref\WittenCR{
E.~Witten,
``Some computations in background independent off-shell string 
theory'',
Phys.\ Rev.\ D {\bf 47}, 3405 (1993)
[hep-th/9210065].
}

\lref\WittenQY{
E.~Witten,
``On background independent open string field theory'',
Phys.\ Rev.\ D {\bf 46}, 5467 (1992)
[hep-th/9208027].
}

\lref\LiZA{
K.~Li and E.~Witten,
``Role of short distance behavior in off-shell open string field theory'',
Phys.\ Rev.\ D {\bf 48}, 853 (1993)
[hep-th/9303067].
}

\lref\GopakumarRW{
R.~Gopakumar, S.~Minwalla and A.~Strominger,
``Symmetry restoration and tachyon condensation in open string theory'',
JHEP {\bf 0104}, 018 (2001)
[hep-th/0007226].
}

\lref\KlebanPF{
M.~Kleban, A.~E.~Lawrence and S.~Shenker,
``Closed strings from nothing'',
[hep-th/0012081].
}

\lref\GerasimovGA{
A.~A.~Gerasimov and S.~L.~Shatashvili,
``Stringy Higgs mechanism and the fate of open strings'',
JHEP {\bf 0101}, 019 (2001)
[hep-th/0011009].
}

\lref\BlumGW{
J.~D.~Blum and K.~R.~Dienes,
``Strong/weak coupling duality relations for non-supersymmetric string  
theories'',
Nucl.\ Phys.\ B {\bf 516}, 83 (1998)
[hep-th/9707160].
}

\lref\BergmanKM{
O.~Bergman and M.~R.~Gaberdiel,
``Dualities of type 0 strings'',
JHEP {\bf 9907}, 022 (1999)
[hep-th/9906055].
}

\lref\CostaNW{
M.~S.~Costa and M.~Gutperle,
``The Kaluza-Klein Melvin solution in M-theory'',
JHEP {\bf 0103}, 027 (2001)
[hep-th/0012072].
}

\lref\FischlerTB{
W.~Fischler and L.~Susskind,
``Dilaton Tadpoles, String Condensates And Scale Invariance. 2'',
Phys.\ Lett.\ B {\bf 173}, 262 (1986).
}

\lref\LambertFN{
N.~D.~Lambert and I.~Sachs,
``String loop corrections to stable non-BPS branes'',
JHEP {\bf 0102}, 018 (2001)
[hep-th/0010045].
}

\lref\SenMG{
A.~Sen,
``Non-BPS states and branes in string theory,''
hep-th/9904207.
}

\lref\SenNX{
A.~Sen and B.~Zwiebach,
``Tachyon condensation in string field theory,''
JHEP {\bf 0003}, 002 (2000)
[hep-th/9912249].
}

\lref\SenMD{
A.~Sen,
``Supersymmetric world-volume action for non-BPS D-branes,''
JHEP {\bf 9910}, 008 (1999)
[hep-th/9909062].
}

\lref\YiHD{
P.~Yi,
``Membranes from five-branes and fundamental strings from Dp branes,''
Nucl.\ Phys.\ B {\bf 550}, 214 (1999)
[hep-th/9901159].
}

\lref\BergmanXF{
O.~Bergman, K.~Hori and P.~Yi,
``Confinement on the brane,''
Nucl.\ Phys.\ B {\bf 580}, 289 (2000)
[hep-th/0002223].
}

\lref\CallanBC{
C.~G.~Callan, C.~Lovelace, C.~R.~Nappi and S.~A.~Yost,
``String Loop Corrections To Beta Functions,''
Nucl.\ Phys.\ B {\bf 288}, 525 (1987).
}

\lref\HughesBW{
J.~Hughes, J.~Liu and J.~Polchinski,
``Virasoro-Shapiro From Wilson,''
Nucl.\ Phys.\ B {\bf 316}, 15 (1989).
}

\lref\MaldacenaRE{
J.~Maldacena,
``The large N limit of superconformal field theories and supergravity,''
Adv.\ Theor.\ Math.\ Phys.\  {\bf 2}, 231 (1998)
[Int.\ J.\ Theor.\ Phys.\  {\bf 38}, 1113 (1998)]
[hep-th/9711200].
}

\lref\WittenQJ{
E.~Witten,
``Anti-de Sitter space and holography'',
Adv.\ Theor.\ Math.\ Phys.\  {\bf 2}, 253 (1998)
[hep-th/9802150].
}

\lref\GubserBC{
S.~S.~Gubser, I.~R.~Klebanov and A.~M.~Polyakov,
``Gauge theory correlators from non-critical string theory,''
Phys.\ Lett.\ B {\bf 428}, 105 (1998)
[hep-th/9802109].
}

\lref\WittenCC{
E.~Witten,
``Noncommutative Geometry And String Field Theory'',
Nucl.\ Phys.\ B {\bf 268}, 253 (1986).
}

\lref\DiVecchiaRH{
P.~Di Vecchia and A.~Liccardo,
``D branes in string theory. I'',
[hep-th/9912161].
}

\lref\EllwoodPY{
I.~Ellwood and W.~Taylor,
``Open string field theory without open strings'',
[hep-th/0103085].
}

\lref\RastelliVB{
L.~Rastelli, A.~Sen and B.~Zwiebach,
``Boundary CFT Construction of D-branes in Vacuum String Field Theory'',
[hep-th/0105168].
}

\lref\KawanoFN{
T.~Kawano and K.~Okuyama,
``Open string fields as matrices'',
[hep-th/0105129].
}

\lref\HarveyNA{
J.~A.~Harvey, D.~Kutasov and E.~J.~Martinec,
``On the relevance of tachyons'',
[hep-th/0003101].
}

\lref\GrossRK{
D.~J.~Gross and W.~Taylor,
``Split string field theory. I'',
[hep-th/0105059].
}

\noindent

\newsec{Introduction}

Tachyon condensation provides a natural arena for the study of off-shell
string theory \refs{\KosteleckyNT,\SenMG,\SenNX}. 
Important intuition about classical 
string field theory has
been developed in this concrete setting. However, 
some of the most
interesting issues raised by the study of open string tachyon 
condensation involve the nature of the quantum theory:
\item{(1)}
What is the significance of the closed strings? 
They appear as poles
in open string loop diagrams and they are the only viable excitations
remaining after the open strings condense into their vacuum state,
 but there is no satisfactory understanding of closed strings in classical
open string field theory.  See 
\refs{\StromingerZD,\SrednickiRV,\HarveyJT,\GibbonsHF,\SenKD,\GerasimovGA,\ChalmersDJ} 
for various points of view.
\item{(2)}
Is the open string vacuum strongly 
coupled? 
The spacetime action suggests
a large effective coupling $\sim 1/V(T)$ near the vacuum at $V(T)=0$
\refs{\SenXM,\YiHD, \SenMD, \BergmanXF,\KlebanPF},
but the world-sheet expansion indicates weak coupling, because holes are
weighted by $\sim V(T)$ \refs{\HarveyNA,\KutasovQP}.
\item{(3)}
Are unstable branes meaningful at finite coupling? 
Duality relations between distinct unstable theories 
have been proposed and could be extremely
important ({\it e.g.} \refs{\BlumGW,\BergmanKM,\CostaNW}), 
but to make them precise one must define such theories
even at large coupling where quantum corrections  dominate.

\noindent
Questions like these force us to confront subtleties of string field theory
that are absent in classical computations.

The most convenient framework for our considerations is boundary string field
theory (BSFT) \refs{\WittenQY,\WittenCR,\ShatashviliKK,\ShatashviliPS,
\GerasimovZP,\KutasovQP}. 
BSFT has so far provided a good understanding of tachyon
condensation at the classical level, based on extending the conformal
sigma model approach to field theories that break 
conformal invariance
on the boundary of a world-sheet with disk topology. The question we shall
address in this paper is how to include quantum corrections in BSFT.

To develop intuition about what corrections can be expected, we first
consider the one-loop correction to the tension of an unstable D-brane without
any world-volume fields excited. This system can be studied in standard 
perturbative string theory, so that we can temporarily avoid the subtleties
associated with going off-shell. The relevant annulus amplitude 
na\"{\i}vely diverges due to the exchange of light closed string states as
well as the open-string tachyon running in the loop. However, these 
divergences can be removed by the Fischler-Susskind mechanism 
\refs{\FischlerTB,\CallanBC}
and analytical
continuation \refs{\MarcusVS}, 
respectively.  The analytic continuation generates an imaginary 
part, which
is interpreted as the decay width of the brane. An amusing qualitative
remark is that the decay of the brane is spatially inhomogeneous, in contrast
to the ``roll down the hill'' usually implied in discussions of tachyon
condensation. These results are in fact quite familiar in other
contexts \refs{\WeinbergVP,\GuthYA}; 
we review them in section 2 because only some of them have previously
been adapted to the discussion of tachyon condensation in string theory
\refs{\MarcusVS}.

One perspective on open string field theory is that it is the generalization
of standard quantum field theory to a situation with infinitely many fields,
one for each open string mode. A natural starting point for quantization is
thus a classical action depending on all the open string modes. However, when
studying quantum corrections in this way, one has to make sure that one really 
starts from the full classical action with all fields included, as opposed to 
an effective action with some of the fields having been integrated out. In
section 3, we discuss this approach to quantum string field theory in the
context of BSFT. We will be left with unresolved issues that render it 
unclear (to us) whether this program can be carried out in BSFT.
If it could, it would seem to lead to a description that is strongly 
coupled around the closed string vacuum.

We therefore seek an alternative determination of loop corrections 
to classical BSFT. Since the classical BSFT action is essentially given by
the partition function on the disk, one might expect that the first loop
correction corresponds to the partition function on a world-sheet of cylinder
topology. However, because the boundary interactions break conformal 
invariance this result would depend on the choice of Weyl factor. In 
section 4 we discuss problems with apparently natural choices of 
world-sheets \refs{\ViswanathanCS,\AlishahihaTG}. 
We are led to the 
physical condition that {\it open strings must couple consistently to closed 
strings even off-shell}. This condition can be satisfied 
by starting with the tree-level couplings to closed strings and
defining loop amplitudes by demanding factorization in the closed string 
channel. A feature of this definition is that it leads to a description 
that is {\it weakly} coupled
around the closed string vacuum.

The paper is organized as follows. In section 2 we perform
the  computation of the on-shell loop amplitude for an
unstable brane and clarify its physical meaning. Then, in section~3, we
review classical BSFT and explore whether a classical action can be
obtained that would be a good starting point for quantization.
Finally, in section 4 we advance our proposal for one-loop BSFT, based
on the requirement of off-shell factorization in the closed
string channel.

\newsec{One Loop Corrections to the D-brane Tension}

In this section we derive the one-loop correction to the tachyon potential  
at the unstable maximum with vanishing tachyon $T=0$; 
in other words, we compute the one-loop correction to the D-brane tension. 
This is an on-shell question which can be answered in detail without
considering the subtleties arising when conformal invariance is broken by 
the tachyon background.

\subsec{Introductory Remarks}
D-branes are solutions of classical string theory with tensions 
proportional to $1/g_s$.  
We want to compute the leading quantum correction to this result. 
Of course, for BPS D-branes all corrections to the tension will 
vanish, so to have a
nontrivial problem we should consider a non-BPS D-brane.  A 
distinction should
be made between those non-BPS D-branes which are classically stable versus 
those which are
classically unstable due to an open string tachyon.  Since our ultimate goal 
is to connect
with issues concerning tachyon condensation we focus on unstable D-branes, in 
particular those
of the bosonic string.   However,  our considerations could also be applied 
to stable
D-branes (for another analysis see \refs{\LambertFN}). 

So we will analyze the question: what is the leading quantum correction to 
the tension
of a bosonic D-brane?  Since these D-branes are unstable one does not expect 
them to exist
as energy eigenstates in the full quantum theory.  As is familiar, when one 
tries to compute
the energy of such an object one finds a complex number, with the imaginary 
part being 
related to the object's decay rate.   This will also turn out to be the case 
here.

The order $(g_s)^0$ correction to the D-brane tension will get contributions
from various sources. Before going into concrete computations, it is useful
to discuss the structure of the expected corrections. Consider a 
$p$-dimensional object coupled to $D$-dimensional gravity
as follows (this should be thought of as part of a low energy effective
action of string theory with a D$p$-brane source):
\eqn\simple{
S={1\over g_s^2}\int d^Dx\sqrt{G}~R\ +\ \tau\int_{p+1}d^{p+1}\xi
\sqrt{\tilde G} ~L_{\rm brane}~,
} 
where $\tilde G$ is the pull-back of the spacetime metric to the D-brane
world-volume.
First work in the limit of small $g_s$ and take the source to have tension $\tau
=\tau^{(-1)}/g_s$ in this limit ($\tau^{(-1)}$ is a constant). The backreaction
of the source on the metric can be ignored in the $g_s\rightarrow 0$ limit, so
we can take $G_{\mu\nu}=\eta_{\mu\nu}$. Now when $g_s$ is non-zero, there
will be quantum corrections to the parameter $\tau$ in \simple\ (corresponding
to loops in the D-brane world-volume theory), 
\eqn\corrtau{
\tau={\tau^{(-1)}\over g_s}+\tau^{(0)}+g_s\tau^{(1)}+\ldots~,
}
and there will be backreaction on the metric,
\eqn\metric{
G_{\mu\nu}=\eta_{\mu\nu}+g_s h_{\mu\nu}^{(1)}+(g_s)^2 h_{\mu\nu}^{(2)}+\ldots~.
}
The total tension of the brane, up to order $(g_s)^0$, will 
schematically look like
\eqn\totaltension{
{\tau^{(-1)}\over g_s}+\tau^{(0)}+\tau^{(-1)}h^{(1)}+(h^{(1)})^2~.
}
The last term in \totaltension\ corresponds to the energy in the gravitational
field excited by the brane. In most of this section we shall concentrate
on the second and third term, which come from the first loop correction 
to the matter energy momentum tensor.

In string theory, one considers a sigma model with background metric \metric\
(and analogously with background values for the dilaton, which we should
also include in this discussion).
Working perturbatively in $g_s$, one can also work with a sigma model in 
Minkowski background and account for the gravitational background by
inserting ``Fischler-Susskind'' (FS) graviton vertex operators in amplitudes.
The classical tension of a D-brane, corresponding to the first term in
\totaltension, stems from the disk without FS 
vertex operators. The second term in \totaltension\ comes from the 
cylinder  (without FS vertex operators), whereas the 
third term arises from the disk with an order $g_s$ FS 
vertex operator. The fourth term corresponds to the 
sphere with two order $g_s$ FS vertex operators.

In the case of a space-filling D-brane, it turns out that the second term in
\totaltension\ has a divergence due to the exchange of light closed string
modes. This divergence is cancelled by the contribution of the disk with an
FS vertex operator. Historically, this is how it was 
discovered that including world-sheets with boundaries leads to modifications
of the classical conformal invariance conditions \refs{\CallanBC}. 
A similar divergence cancellation mechanism was earlier 
discovered by Fischler and Susskind \refs{\FischlerTB}. From a 
modern perspective, 
the shift in the sigma model metric is nothing but the 
backreaction of a D-brane on the space-time geometry. 

\subsec{The Cylinder Amplitude}
For definiteness we consider a spacefilling D25-brane in bosonic string 
theory.   The leading 
order tension is equal to the disk partition function, with the result 
$\tau \sim (\alpha')^{-13} / g_s$.
The vacuum amplitude on the cylinder is given by the well known expression
\eqn\b{Z_{{\rm cyl}} = iV_{26} \int^{\infty}_{0} \! {dt \over 2t}\, 
(8 \pi^2 \alpha' t)^{-13} 
\eta(it)^{-24} = {iV_{26} \over 2\pi (8 \pi^2 \alpha')^{13}} \int^{\infty}_0 
\! ds \, \eta(is/\pi)^{-24}~,}
where
\eqn\c{\eta(ix)^{-24} = \exp(2\pi x ) \prod_{n=1}^{\infty}
\left[1-\exp(-2\pi nx)\right]^{-24}
= \exp(2\pi x) + 24 + O(\exp(-2\pi x))~.} 
Recall that large $t$ and large $s$ correspond to the limits of short and 
long cylinders 
respectively.  The cylinder amplitude diverges in both of these limits; the 
large $t$ divergence
is due to the open string tachyon, while the large $s$ divergence is due to 
the closed string
tachyon, dilaton and graviton.   These two divergences are rendered finite by 
different 
mechanisms, which we now review.

We begin with the large $s$ divergence.  Imposing a 
cutoff $s= \ln{\Lambda_s}$ on the upper range of the $s$
integration, the divergent contributions are
\eqn\cz{Z_{{\rm cyl}}^{\rm div} = {iV_{26} \over 2\pi (8 \pi^2 \alpha')^{13}} \left(
{1 \over 2} \Lambda_s^2 + 24 \ln{\Lambda_s} \right)~.}
The divergences arise because the D-brane acts as a zero momentum source for 
the closed string 
tachyon, dilaton, and graviton.   The self energy correction then involves 
evaluating propagators
for these fields at zero momentum, which yields a divergent result in the 
proper time representation:
\eqn\d{\left({1 \over k^2 + m^2}\right)_{k^2=0} \longrightarrow 
{\alpha' \over 2} \int_0^{\ln{\Lambda_s}} \! 
ds \, e^{-{1 \over 2}\alpha' m^2 s} ~~\sim ~~
\left\{\eqalign{  ~~ \Lambda_s^2~, ~~~~ \alpha' m^2 = &-4  \cr  \ln{\Lambda_s}~, 
~~~ \alpha' m^2 &=0~.   } \right.}

Since we are ultimately more interested in the superstring,
 we will focus on the $\log$ divergence due to massless closed string exchange.  
In  field theory one finds the same divergences when one tries to expand around 
a constant field configuration that is not a solution to the equations of motion, 
and the resolution is simply that one should shift the field to an extremum of 
the potential.   The string theory analogue is the Fischler-Susskind mechanism, 
which instructs one to cancel the divergences by evaluating amplitudes in a nontrivial 
closed string background.   Thus on the disk one should add a vertex operator proportional 
to  $h_{\mu\nu} \partial X^\mu \bar{\partial}X^\nu$.  Here 
$h_{\mu\nu} \sim g_s \ln\Lambda_s \eta_{\mu\nu}$  is the background produced by a constant 
source for the dilaton and graviton, and is logarithmically divergent as above due to 
the zero momentum singularity of the propagator.   If one equates $\Lambda_s$ with the 
world-sheet cutoff then one can think of the Fischler-Susskind mechanism as yielding a 
loop correction to the world-sheet beta functions, or equivalently, as adding source
terms to the dilaton and graviton equations of motion.  Similarly, the
quadratic divergence is removed by shifting the closed string tachyon.

So the full amplitude at order $(g_s)^0$ is obtained by adding together the cylinder, 
the disk with the insertion of one closed string vertex operator, and the sphere 
with the insertion of two closed string vertex operators. The sum is what one would 
measure as the correction to the D-brane tension (see \totaltension). In fact, for a spacefilling D-brane 
the situation is more complicated.  The actual closed string 
background would involve a rolling dilaton that becomes large somewhere in spacetime, thus 
invalidating our perturbation theory built around flat spacetime with constant weak string
coupling. Symptomatic of this is the fact that our sphere contribution is ill-defined due 
to the divergent coefficient of the closed string vertex operator. Keeping this in mind, 
we will in the following just focus on the open string contribution; that is, the cylinder 
contribution with  divergences cancelled against the disk as above. It is this quantity --- 
which essentially gives the zero point energy of open strings on the D-brane --- that 
is the closest analogue of the standard field theory version of the one loop effective 
potential.  In any event, any realistic computation of, say, the tension of a 
non-spacefilling D-brane, will involve computing the analogue of this quantity.

%

\subsec{Interpretation of the Open String Tachyon Divergence}
At this stage our cylinder amplitude has been rendered finite in the large $s$ limit and
reads
\eqn\e{\eqalign{ \tilde{Z}_{{\rm cyl}} = Z _{{\rm cyl}} - Z_{{\rm cyl}}^{{\rm div}} 
 &= {iV_{26}  \over 2\pi (8 \pi^2 \alpha')^{13}} \int^{\infty}_0 \! ds \, \left\{\eta(is/\pi)^{-24}
- \exp(2s) - 24 \right\} \cr
&= {iV_{26} \over (8\pi^2 \alpha')^{13}} \int^{\infty}_{0} \! {dt \over 2t}\, 
\left\{ {\eta(it)^{-24} \over t^{13}}   - {\exp(2\pi/t) \over t}  - {24 \over t} \right\}~.
}}
This expression, with the closed string divergences subtracted, is the one-loop
correction to the tension according to the terminology introduced above.
It is finite in the $t \rightarrow 0$ limit by construction but diverges in 
the $t \rightarrow \infty$ limit due to the behavior \c\ of $\eta(it)^{-24}$.
This divergence is due to the open string tachyon. To make sense of this, consider as 
an example the one loop vacuum-to-vacuum amplitude for 
a point particle of mass $m$,
\eqn\f{Z(m^2) = i V_{26} \int_0^\infty \! {d\ell \over 2 \ell} \int {d^{26}k \over (2\pi)^{26}}
e^{-(k^2+m^2)\ell/2} = {i V_{26} \over (2\pi)^{13}} \int_0^\infty \! {d\ell \over 2 \ell}\,
{e^{-m^2 \ell/2} \over \ell^{13}}~.
}
Inserting the open string tachyon mass $\alpha' m^2 = -1$ and identifying $\ell = 4\pi \alpha' t$,
we find the same large $t$ divergence as in \e.  More generally, the one loop Coleman-Weinberg
effective potential is obtained from \f\ by setting $m^2 = V''(\phi)$ where $V(\phi)$ is the
tree level potential; the effective potential diverges whenever the tree level potential
satisfies $V'' < 0$.   There exists a good understanding of this divergence, 
and a well defined prescription
for rendering it finite, which we can apply in the string theory context.

To understand the issues involved,  recall the physical meaning of the effective 
potential.  Consider a scalar field theory with tree level potential $V(\phi)$.   Let
$\left\{|\Psi_{\bar{\phi}} \rangle\right\}$ denote the space of quantum states satisfying 
\eqn\g{\langle \Psi_{\bar{\phi}} | \hat{\phi}(x) |\Psi_{\bar{\phi}} \rangle = \bar{\phi} = 
{\rm constant}~.} 
Now minimize the expectation value of the Hamiltonian in this space of states to obtain 
$E(\bar{\phi})$,
assuming that such a minimum exists.   Then, $V_{{\rm eff}}(\phi)= E(\phi)$ is the 
effective potential. From 
this definition, it is straightforward to show that the effective potential is convex,
$V''_{{\rm eff}}(\phi) \geq 0$, which implies that $V(\phi)$ is a poor approximation to 
$V_{{\rm eff}}(\phi)$ whenever $V''(\phi)<0$.   
For example, choose a $\bar{\phi}$ in a region between two adjacent minima of $V(\phi)$;
then the state with minimum energy is not one concentrated near $\bar{\phi}$, but 
rather a linear combination of states localized near the minima.   

It is now immediately clear that the effective potential for a tachyonic 
mass scalar
field is ill-defined for any value of $\phi$, since the energy in the space of  states with a 
fixed expectation
value of $\hat{\phi}$ is unbounded from below.   This explains the divergence in \e\ and \f.  It
is also now clear that to obtain a finite result we should ask a more physically well defined 
question. 
A more reasonable question to ask is what is the energy of a state well localized near a given 
value of
$\phi$.  Note however that the answer to this question will depend on which particular state is 
chosen; furthermore, since no such state is an eigenstate of the Hamiltonian one can expect the
energy to acquire an imaginary part:
\eqn\h{ \langle \Psi | e^{-i H t}|\Psi\rangle  = e^{- {\rm Im}(E)t - i {\rm Re}(E)t}~,}
for large $t$.  It turns out that ${\rm Im}(E)$ is largely independent of the particular state
chosen, and that for a particular ``natural'' choice of the state one can obtain both the real
and imaginary parts by analytically continuing our original formula \f.  This particular state
corresponds to placing each Fourier mode  $\phi_k$ in a minimum uncertainty wavepacket 
concentrated at the origin, except for the zero mode which is taken to be an eigenstate at
the desired value of $\phi$.   
We refer to \WeinbergVP\ for more details.  

To analytically continue \f, first divide the integration region into two parts (see \MarcusVS\ for
a similar computation in a related context)
\eqn\ieq{I(m^2) = \int_0^\infty \! {d\ell \over 2 \ell}\,
{e^{-m^2 \ell/2} \over \ell^{13}}  = \int_0^{\hat{\ell}} \! {d\ell \over 2 \ell}\,
{e^{-m^2 \ell/2} \over \ell^{13}} +
\int_{\hat{\ell}}^\infty \! {d\ell \over 2 \ell}\,
{e^{-m^2 \ell/2} \over \ell^{13}}~.}
The interesting divergence is in the second term, which we rewrite as
\eqn\j{\left({m^2}\right)^{13} \int_{m^2 \hat{\ell}}^\infty \! {dx \over 2 x}\,
{e^{- x/2} \over x^{13}}.}
Now continue \j\ to $m^2 <0$ by taking an integration contour in the upper half plane
 that avoids the singularity at the
origin.    The analytic continuation yields an imaginary part for $m^2 <0$,
\eqn\k{i{\rm Im}\left(I(m^2) \right) = - {1 \over 2}
(m^2)^{13} \oint_C \! {dz \over 2 z}\,
{e^{- z/2} \over z^{13}} = - {i \pi \over 2 \cdot 13 !} \left( {-m^2 \over 2}\right)^{13}~.
}
where $C$ is a contour circling the origin in the clockwise direction.
Then from \f\ this translates into an imaginary part for the energy given by
\eqn\l{ {\rm Im}(E) = - {V_{25} \over (2\pi)^{13}} {\rm Im}\left( I(m^2)\right) 
= {\pi V_{25} \over 2 \cdot 13!} \left({-m^2 \over 4 \pi}\right)^{13}.}
The decay rate of the system is given as usual by $\Gamma = 2{\rm Im}(E)$.    The real part of 
$I(m^2)$ can also be extracted by evaluating the integral in the complex plane. It is easy
to check that the result is independent of our choice for $\hat{\ell}$.  

We can use precisely the same strategy to define our integral \e.
Dividing up the range of 
integration, and using the asymptotic behavior \c, we can write \e\ as
\eqn\m{\tilde{Z}_{{\rm cyl}} = {iV_{26} \over (8\pi^2 \alpha')^{13}} \int_{\hat{t}}^{\infty} 
\! {dt \over 2t}\,
{\exp(2\pi t) \over t^{13}} +  \tilde{Z}_{{\rm cyl}}^{{\rm finite}}~.}
$\tilde{Z}_{{\rm cyl}}^{{\rm finite}}$ is a well defined 
and purely imaginary expression 
given explicitly
by
\eqn\n{\eqalign{\tilde{Z}_{{\rm cyl}}^{{\rm finite}}=&
{iV_{26} \over (8\pi^2 \alpha')^{13}} \int^{\hat{t}}_{0} \! {dt \over 2t}\, 
\left\{ {\eta(it)^{-24} \over t^{13}}   - {\exp(2\pi/t) \over t}  - {24 \over t} \right\} \cr
& + {iV_{26} \over (8\pi^2 \alpha')^{13}} \int_{\hat{t}}^{\infty} \! {dt \over 2t}\, 
\left\{ {\eta(it)^{-24}- \exp(2\pi t) \over t^{13}}   - {\exp(2\pi/t) \over t}  - 
{24 \over t} \right\}~.}}
Next, we analytically continue in the open string tachyon mass exactly as before to obtain
\eqn\o{\tilde{Z}_{{\rm cyl}} = -{iV_{26} \over (8\pi^2 \alpha')^{13}} \int_C \! {dz \over 2z}\,
{\exp(-2\pi z) \over z^{13}} +  \tilde{Z}_{{\rm cyl}}^{{\rm finite}},}
where $C$ is a contour running from $z=-\hat{t}$ to $z=\infty$ and avoiding the singularity in the
origin; by convention we take $C$ to lie in the upper half complex plane, as 
 shown in Fig.~1.  
\fig{Contour used in evaluating \o.}{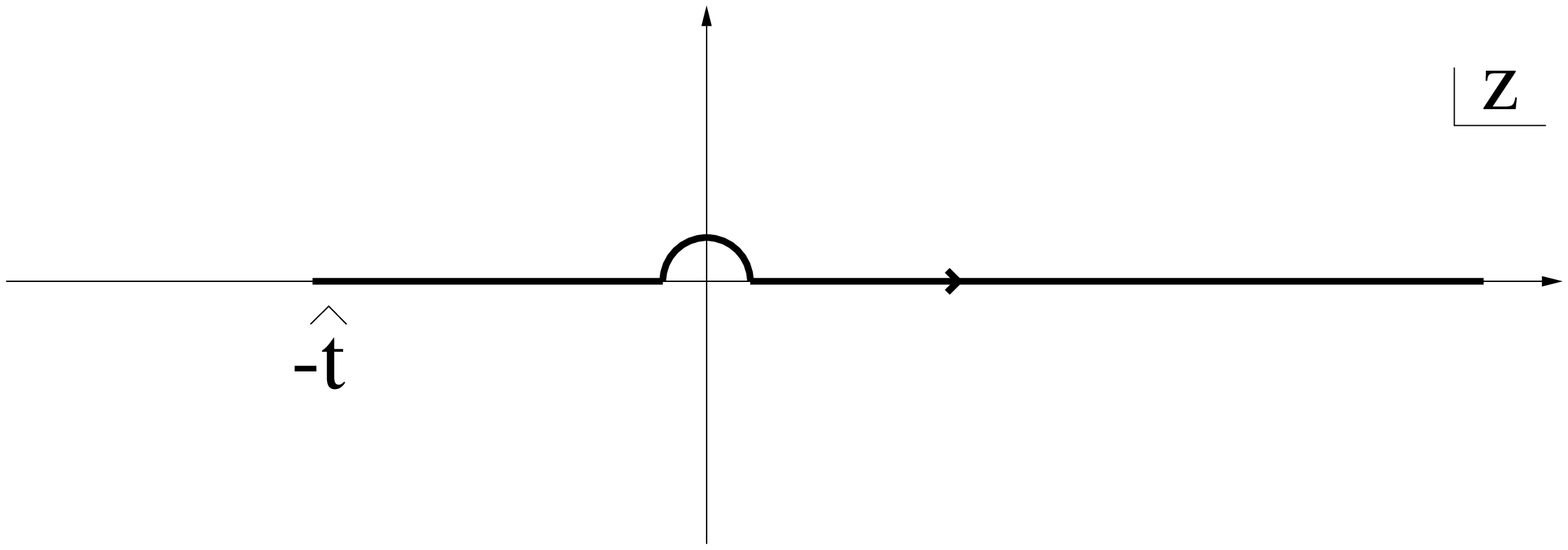}{4.5truein}     

Therefore,
\eqn\p{{\rm Im}(i\tilde{Z}_{{\rm cyl}}) = 
{ \pi V_{26} \over 2 \cdot 13!\cdot (4\pi \alpha')^{13}}~,}
and
\eqn\q{{\rm Im}(E) = { \pi V_{25} \over 2 \cdot 13!\cdot (4\pi \alpha')^{13}}~.}
The contribution to the real part of $E$ can also be extracted from \o.  

Although we obtained ${\rm Im}(E)$ by a process of analytic continuation, we should again stress
that it has a precise physical meaning in terms of the decay rate of an initial state localized
at the unstable point of the tachyon potential.  In the string theory context \q\  gives the 
leading order decay rate of a bosonic D25-brane.   We also note that although we found it 
convenient to compute the decay rate in terms of the imaginary part of a loop amplitude, it is 
essentially a tree level quantity in the sense that it {\it could} have been obtained by 
solving the tree level equations of motion. 

Let us now say a few words about the manner in  which the initial state decays.  One's initial
intuition is that the D25-brane  decays via the tachyon rolling homogeneously down the
potential.   However, the actual decay process is inhomogeneous and proceeds by the amplification
of long wavelength modes of the tachyon (spinodal decomposition).  
In infinite volume the zero mode of the tachyon is a classical variable and can be 
taken to be fixed at the top of the potential.  Modes of the tachyon obeying $\vec{k}^2 + m^2 < 0$
are unstable, and grow exponentially in time, with the longest wavelengths growing most rapidly.
Thus there will be large regions of space in which the tachyon rolls down the left hand side of
the potential, and large regions in which it rolls down the right hand side.  In the bosonic 
string the potential seems to be unbounded from below in one direction, and so the actual outcome
of the decay process is uncertain.  For the superstring the potential is bounded from below,
and the decay will eventually terminate in a distribution of topological defects.

\newsec{Classical and Quantum BSFT}
We would like to generalize the computations in the previous section to
allow off-shell open strings, such as a tachyon condensate. In this 
section we review BSFT and attempt to define loop corrections in BSFT by quantizing
the tree level action as an ordinary field theory.

\subsec{Classical BSFT}

Boundary string field theory (BSFT) describes the off-shell dynamics of open 
strings in a fixed on-shell background of  closed strings. An open  string 
field configuration 
corresponds to a boundary term in the world-sheet action of the string. 
Specifying a boundary term means giving the background values of the various
modes of the open string. In the case of bosonic open string 
theory, to which we specialize now, these 
modes are tachyons, gauge potentials and massive string states. 
BSFT formally defines an action functional on this space of string field
configurations, equal to the {\it renormalized} 
partition function 
on the unit disk of the world-sheet theory, 
up to a correction involving the world-sheet
beta functions.

In formulas, one starts with a world-sheet theory 
\eqn\bone{
{\cal S}= {\cal S}_0+\int_0^{2\pi}{d\tau\over 2\pi}{\cal V}~,
}
where ${\cal S}_0$ is conformally invariant and the boundary perturbation 
${\cal V}$ can be expanded as
\eqn\btwo{
{\cal V}=\sum_i\lambda^i{\cal V}_i~.
}
The world-sheet couplings correspond to the modes of spacetime fields. In 
terms of the $\beta$-functions for these world-sheet couplings, the 
classical space-time action $S(\lambda)$ is defined 
as \refs{\WittenCR,\ShatashviliPS}
\eqn\bthree{
S=(\beta^i{\partial\over\partial\lambda^i}+1)\,Z(\lambda)~,
}
where $Z$ is the renormalized disk partition function of the theory defined
by \bone. In fact, the 
$\beta$-function term in \bthree\ subtracts the divergence due to the infinite
volume of the M\"obius group, and could be viewed as part of the 
renormalization procedure (it corresponds to a singular wave function 
renormalization of the tachyon). It also ensures 
that $S$ has extrema when the 
boundary theory is conformal
\foot{In the case of the superstring the BSFT action simply equals the 
renormalized disk partition function without 
$\beta$-function corrections 
\refs{\KutasovAQ,\KrausNJ,\TakayanagiRZ,\MarinoQC,\NiarchosSI}.}.
Note that for a conformally invariant boundary
action ($\beta^i=0$) the spacetime action equals the disk partition function.

For later reference we note 
that another way of writing \bthree\ is \refs{\ShatashviliPS,\KutasovAQ}
\eqn\bthreebis{
{\partial S\over\partial\lambda^i}=\beta^jG_{ij}(\lambda)~,
}
where $G_{ij}$ is the metric on the space of theories (we shall not need its
explicit form in this paper).

The fact that the renormalized disk partition function can be identified with 
the spacetime action is familiar from the sigma model approach to string 
theory \refs{\FradkinPQ} (for reviews see {\it e.g.} \refs{\TseytlinRR,
\TseytlinMT, \TseytlinMV}). 
In that approach, most attention has gone to on-shell, renormalizable
boundary interactions, and usually one constructs the action in a derivative
expansion. It is convenient to work in Euclidean spacetime signature, because
then the world-sheet theory is unitary.
Almost by construction, the bare (unrenormalized)
disk partition function is the generating functional of S-matrix elements
(up to a factor of the M\"obius volume that has to be divided out). 
Divergences arise due to the exchange of zero momentum massless open 
string modes and of tachyonic modes with low enough momentum, as is clear 
from the Schwinger parameterization of propagators,
\eqn\schpar{
(p^2+m^2)^{-1} = \int_0^\infty \! ds\, e^{-(p^2+m^2)s}~.
}
 From the world-sheet point of view these are short 
distance divergences  and are removed by renormalizing the partition 
function. Thus, at least in Euclidean signature, the renormalized partition
function  generates diagrams that are one-particle irreducible (1PI) 
with respect to those massless and tachyonic lines. The terms containing only
these renormalizable modes correspond to their effective action with the
unrenormalizable modes integrated out. It has been checked in various
examples that, for massless external open string states and in a derivative
expansion, the S-matrix of perturbative string theory can be reproduced 
from this effective action \refs{\TseytlinRR, \AndreevCB, \TseytlinMT}.  
One just computes the Feynman diagrams corresponding to the
exchanges of the massless and tachyonic modes that were subtracted out.
Some checks have also been performed for tachyonic external states \FrolovNB. 

BSFT is an extension of the sigma model approach which is 
non-trivial in several ways. First, one wants to be able to discuss off-shell
open strings (the closed string background remains on-shell). Further, one 
would like to describe all open string modes, including the ones corresponding 
to irrelevant (unrenormalizable) boundary interactions. And finally, one would 
like to go beyond the low energy expansion. These complications are 
challenging, 
and at present it is unclear whether the ambitions of BSFT can be 
realized even at the classical level. In particular it seems to require 
additional information to treat the irrelevant interactions exactly but 
it is possible to incorporate them perturbatively by expanding in these 
operators.

As discussed above, a conventional renormalization prescription in 
Euclidean signature 
produces a sigma-model action with all irrelevant open string modes 
integrated out classically.
This action is inappropriate away from the classical and low energy limits. 
Instead, what 
one wants is an action that generates 1PI diagrams with respect to all 
modes, whether 
massless, tachyonic, or massive. This would be the sort of action that 
deserves to be 
called a string field theory; for instance, the much studied cubic 
bosonic open string 
field theory \refs{\WittenCC} is of this nature (see \refs{\EllwoodPY,
\RastelliVB, \GrossRK, \KawanoFN} and references therein for some recent
developments). Given such an action, off-shell 
amplitudes for arbitrary 
external states are computed from Feynman diagrams as in an ordinary field
theory.  It may also be necessary to add new vertices at each order in 
perturbation theory in order to recover the correct on-shell loop amplitudes. 

\subsec{Quantum BSFT}
Let us summarize the situation: tree level BSFT leads to a spacetime ``action'' 
with the irrelevant modes integrated out classically. Unfortunately such an action 
is not an acceptable starting point for including loop corrections; for instance, the 
effects of massive states running in loops cannot be taken into account if they have 
been integrated out classically. 

At this point one might simply give up on defining loop amplitudes
starting from tree-level BSFT, concluding that BSFT necessarily 
is an effective field theory description limited to the light open 
string modes. This pessimistic view is consistent with the fact
that the successes of BSFT so far only probe relevant and marginal 
operators at the classical level. In the remainder of this section
we discuss the alternative and more optimistic view that a more 
fundamental classical action can be defined which describes all open
string modes and generates truly 1PI diagrams.
 
What we need is a different renormalization prescription that subtracts, from
the spacetime point of view, divergences due to the exchange of any on-shell
open string state. Such a prescription would seem to arise most naturally
in Lorentzian signature where the regularized
propagator takes the form
\eqn\schw{{1 \over p^2-m^2} = -\int_0^{\ln \Lambda} \! ds \, 
e^{(p^2-m^2)s}~.}
Subtracting the logarithmic divergences will then remove the exchange of
any on-shell state. This prescription is in fact not entirely well-defined 
as it stands because now the underlying CFT is non-unitary; however,
it serves to indicate what is needed. In the following discussion we will 
only rely on a few structural features of the action. In particular, 
we will not need a precise definition of our renormalization prescription 
beyond saying that it should subtract divergences from all on-shell 
exchanges.  

According to \bthreebis, the equations of motion are $\beta_i=0$ where
$i$ enumerates a complete set of operators. These are not the familiar
sigma-model conditions for conformal invariance: to recover 
such standard results as the Born-Infeld action, one would need to 
eliminate the massive fields by solving their equation of motion. 
Banks and Martinec have defined a string field theory with these 
``complete'' equations of motion \refs{\BanksQS} and it was shown 
that they imply the correct tree level 
S-matrix amplitudes \refs{\HughesBW}. Their 
$\beta_i$ actually refer to the Wilsonian beta functions but the 
relation between Wilsonian and ``conventional'' beta functions 
corresponds to a redefinition of couplings so one expects the same 
success in our construction. 

We can make the action slightly more explicit by expanding it 
around the classical solution $\lambda_i=0$ 
\eqn\expand{ S = \sum_{n=2}^\infty S_{i_1 \cdots i_n} \lambda^{i_1} \cdots
\lambda^{i_n}~.} For constant tachyon background  
$\int \! {d\tau \over 2\pi} \,T = T$ we have
$Z(T,\lambda^i) = e^{-T}Z(0,\lambda^i)$ so, after taking \bthree\ into 
account, the expansion \expand\ takes the form
\eqn\expandb{S = \sum_{n=2}^\infty (a_{i_1 \cdots i_n}+b_{i_1 \cdots i_n}T)
e^{-T}  \lambda^{i_1} \cdots
\lambda^{i_n}~,}
where now the $\lambda^i$ exclude $T$.

We can try to compute loop amplitudes using the action \expandb . 
By construction, the loop amplitudes obtained this way will factorize 
correctly in the open string channel, and it seems plausible that on-shell 
we will recover the correct S-matrix amplitudes \foot{It would of course 
be worthwhile to check this explicitly as was done in the cubic theory 
\Freedman .}. However, the open string loop amplitudes should also factorize 
properly in the closed string channel and it is far from obvious how this 
is supposed to come about.\foot{See \ZwiebachQJ\ for an alternative string
field theory that makes
factorization manifest, and which bears resemblance to our approach in 
section 4.} Moreover, the theory described by \expandb\ 
apparently becomes strongly coupled at $T\rightarrow\infty$, which is the 
location of the closed string vacuum. This is puzzling, because in the next 
section we will define loop amplitudes that automatically factorize correctly 
in the closed string channel but are weakly coupled at the closed string vacuum. 
It will be an interesting question to find the precise relation between these 
two descriptions.

\newsec{Loop Corrections to the Tachyon Action}
The purpose of this section is to propose a prescription for loop 
computations in BSFT that circumvents the problems discussed in section 3. 
We first explain  the problems with a na\"{\i}ve annulus prescription.
Then the BSFT couplings between open and closed strings are
discussed. Finally, we combine the ingredients and find a factorization 
condition that defines loop amplitudes.
  
\subsec{The Partition Function on the Annulus?}  
The classical BSFT action arises from evaluating the disk partition 
function with the boundary interaction
\eqn\bpdef{
S_{\rm bndy}=\int\! {d\tau\over 2\pi} ~T(X(\tau))~.
}
To define a loop correction one might propose simply adding an additional 
boundary to the disk (weighted by $e^{-S_{\rm bndy}}$) and then integrating
over moduli.  This of course would be in direct analogy to how one computes 
loop corrections to on-shell quantities.  

In  the on-shell case one has a freedom in the choice of a world-sheet 
Weyl factor, and it is customary to use this freedom to let the world-sheet
have the geometry  of either  a cylinder or an annulus. Conformal 
invariance guarantees that on-shell amplitudes are independent of this 
choice. But once we break conformal invariance these world-sheets no 
longer give equivalent results, and one is left with an ambiguity as 
to the choice of Weyl factor. This ambiguity did not
arise at the level of the disk because the definition of BSFT demands
that the Weyl factor be rotationally invariant, and the remaining 
freedom can be compensated for by a redefinition of couplings.  
One possible approach is to contemplate integrating over Weyl factors,
but this seems challenging both technically and conceptually, and also
seems against the spirit of tree level BSFT which assumes a fixed Weyl 
factor.  

To gain a more physical understanding of the issues involved, let us 
see what goes wrong when we simply decree that the annulus is the 
preferred world-sheet.  We take our annulus to have a unit outer radius
and an inner radius $a$.  For Neumann boundary conditions (no boundary
interaction) the annulus amplitude is 
\eqn\zann{Z_{{\rm ann}} = \int_0^1 \! {da \over a^3} \prod_{n=1}^{\infty}
(1-a^{2n})^{-24}~.}
Now include a constant tachyon $T$.  This simply introduces a factor
\eqn\Tbndy{e^{-S_{\rm bndy}} = e^{-(1+a)T}~,}
so the annulus amplitude becomes
\eqn\zannT{Z_{{\rm ann}}(T) = \int_0^1 \! {da \over a^3} \prod_{n=1}^{\infty}
(1-a^{2n})^{-24}e^{-(1+a)T}~.}
The annulus amplitude diverges in the small $a$ region due to the 
exchange of tachyonic and massless closed string modes,
\eqn\Zexp{Z_{{\rm ann}}(T) = \int_0^1 \! {da \over a^3}e^{-T}\left(1 -a T
+ (24 + {1\over 2}T^2) a^2 + O(a^3) \right)~.}
In particular, cutting off the lower limit of integration we find a 
logarithmic divergence equal to
\eqn\logdiv{- (24+{1\over 2}T^2) e^{-T} \log a_{{\rm min}}~.}
In the discussion of the on-shell loop-amplitude in section 2 we
encountered this divergence at $T=0$ and removed it using the 
Fischler-Susskind mechanism; {\it i.e.} it was attributed to the tree 
level exchange of dilatons and gravitons. The amplitude for graviton 
exchange is of the form
\eqn\gravex{   \left. 
{T^{\mu\nu}T_{\mu\nu} \over k^2}\right|_{k^2=0} \rightarrow
T^{\mu\nu}T_{\mu\nu} \log a_{{\rm min}}~,}
and from tree level BSFT we know that the energy momentum tensor in 
the presence of a constant tachyon is proportional to the tachyon potential,
$T_{\mu\nu} \sim (1+T)e^{-T} \eta_{\mu\nu}$. The dilaton 
contribution is of similar form. So we find a 
discrepancy with \logdiv\ : {\it the annulus divergence cannot be 
interpreted as due to exchange of gravitons} that couple to the energy 
momentum tensor of the open strings. It seems unlikely that one could 
develop a consistent formalism in which gravity does not couple to the 
energy momentum tensor, and so we conclude that annular world-sheets 
leads to inconsistencies. 

The most obvious problem with the annulus is that it treats the two
boundaries asymmetrically, while closed string factorization implies
that the two boundaries should be on an equal footing. This immediately
suggests that cylindrical world-sheets are better candidates. In fact,
we will argue that the cylinder also fails to factorize correctly, but that
the discrepancies only arise for nonconstant tachyon backgrounds.  

Rather than making an {\it ad hoc} choice of world-sheet and then checking
for factorization, we will take factorization as our starting point and
derive loop amplitudes based on this physical condition.   
This is a powerful principle because an amplitude with 
any number of boundaries 
(but no handles) can
be represented as a tree level closed string process.   We will concentrate
on the case of two boundaries, which requires us to know the closed string
propagator as well as the vertex coupling open strings to a single closed 
string.

\subsec{Tree Level Couplings to Closed Strings}

We first determine the vertices by evaluating the disk amplitude with 
insertion of a closed string vertex operator.  One is accustomed to 
using   conformal invariance to freely choose the position of the 
closed string vertex operator, or to use an integrated vertex operator.
But in the presence of a non-conformally invariant boundary interaction, 
or for an off-shell closed string vertex operator, these choices are
inequivalent. The correct procedure is to use integrated vertex 
operators. (This certainly seems like the most symmetrical choice and 
arises from a sum over surfaces approach).

To show this, consider the disk partition function in the absence of 
a closed string vertex operator,
\eqn\diskamp{Z = \int \! DX\, e^{-S_{{\rm bulk}} - S_{{\rm bndy}}}~,}
with 
\eqn\sbulk{S_{{\rm bulk}} = {1 \over 4\pi \alpha'}\int \!d^2 x \,
\sqrt{\gamma}\gamma^{ab}G_{\mu\nu}\partial_a X^\mu \partial_b X^\nu~,}
where we take $G_{\mu\nu}=\eta_{\mu\nu}$.
Taking a general tachyon boundary interaction $T(X)$, we will find
\eqn\diskder{Z[T(X)]= Z[T,\eta^{\mu\nu}\partial_{\mu}T \partial_\nu T, \ldots]~.
}
Now consider the vertex
operator corresponding to a zero momentum graviton,  
$V \sim  h_{\mu\nu}\partial X^\mu \bar{\partial}X^\nu$.  Such a zero momentum
graviton is of course just a coordinate transformation, and the effect on
the disk amplitude \diskder\ should just be to replace
$\eta_{\mu\nu} \rightarrow \eta_{\mu\nu} +  h_{\mu\nu}.$   This will
be satisfied provided that the graviton vertex operator is defined as the
functional derivative of the bulk action with respect to the metric; {\it i.e.}
\eqn\vertex{
V = {\delta S_{{\rm bulk}} \over \delta  G_{\mu\nu} }  h_{\mu\nu} 
 \sim \int\! d^2x  \,
\sqrt{\gamma}\gamma^{ab} h_{\mu\nu} \partial_a X^\mu \partial_b 
X^\nu~.}
So general covariance demands that we choose an integrated vertex operator
for the graviton.  Choosing instead, say, a vertex operator fixed at the
origin, one will find that the graviton couples not to the standard energy
momentum tensor but to something else, and this will in all likelihood 
lead to an inconsistency.
 Similarly, we expect that the full set of closed string
gauge invariances require one to use integrated vertex operators in general,
and we will assume this from now on.  

So our vertex coupling a single closed string state with momentum $p$
 to any number of open string is given by evaluating
\eqn\vdef{Z_{{\rm disk}}(V_I(p); S_{{\rm bndy}}) = 
\int \! DX \, e^{-S_{{\rm bulk}} - S_{{\rm bndy}}} V_I(p)~,}
where $V_I(p)$ is an integrated vertex operator for the closed
string state $I$.  In evaluating \vdef\ one will encounter divergences
when the vertex operator approaches the boundary of the disk.  Furthermore,
to obtain the proper normalization of \vdef\ one should divide by the 
volume of the M\"obius group, but this is also divergent.  It was shown
in \LiuNZ\ how to renormalize these divergences. In the on-shell case
the result is the same as if one had fixed the vertex operator at the
origin and divided by the (finite) volume of the unbroken subgroup of the
M\"obius group. 

\subsec{Loops via Factorization}
To construct a ``cylinder amplitude'' using the above vertices we first factorize 
the standard cylinder amplitude (with a conformally invariant boundary interaction) 
in terms of disk amplitudes with closed string vertex operators fixed at the origin:
\eqn\factor{Z_{{\rm cyl}}(S_{{\rm bndy}}) = 
\sum_{I} \int \! {d^{26} p \over (2\pi)^{26}}\,
Z_{{\rm disk}}(V^{{\rm fixed}}_I(p); S_{{\rm bndy}}){1 \over p^2 + m_I^2}
Z_{{\rm disk}}(V^{{\rm fixed}}_I(-p); S_{{\rm bndy}})~,}
where $I$ enumerates all closed string states. One way of deriving this formula
is in the boundary state formalism, see \refs{\DiVecchiaRH} for a review. 
If we simply defined 
off-shell amplitudes by allowing nonconformal $S_{\rm bndy}$ in \factor\ 
then factorization would imply that the couplings to closed strings
were given by the disk amplitude with fixed vertex operators inserted;
but those couplings would be incorrect since we showed above that one 
must use integrated vertex operators.

To correct for this we need to relate the disk amplitudes with fixed and
integrated vertex operators.  Our integrated vertex operators are integrals
over the disk of conformal tensors of dimension 
$(\Delta,\Delta)$ with $ \Delta = 1 + \alpha'(p^2 + m_I^2)/4$.  We can perform
an $SL(2,R)$ transformation to bring any point in the interior of the disk
to the origin.  If the boundary interaction is conformally invariant we
can use this to transform a disk amplitude with an integrated vertex operator
into one with a fixed vertex operator, picking up in the process a function
of $\Delta -1$; so for conformally invariant boundary interactions
\eqn\transf{Z_{{\rm disk}}(V^{{\rm fixed}}_I(p); S_{{\rm bndy}}) =
f(p^2 + m_I^2) Z_{{\rm disk}}(V_I(p); S_{{\rm bndy}})~.}
Since fixed and integrated vertex operators are equivalent onshell
we have $f(0)=1$. Using \transf\ in \factor\ we find
\eqn\factorb{Z_{{\rm cyl}}(S_{{\rm bndy}}) = 
\sum_{I} \int \! {d^{26} p \over (2\pi)^{26}}\,
Z_{{\rm disk}}(V_I(p); S_{{\rm bndy}}){f(p^2+m_I^2)^2  \over p^2 + m_I^2}
Z_{{\rm disk}}(V_I(-p); S_{{\rm bndy}})~.}
{\it We define general off-shell loop-amplitudes by allowing arbitrary
$S_{\rm bndy}$ in this formula.}
This definition has all the properties we desire:  by construction it 
reproduces the correct on-shell amplitudes and also factorizes correctly
in the closed string channel.  In particular, the amplitude for graviton
exchange is given by the correct energy momentum tensor of the open strings.
We stress again that our result is not the same as simply evaluating the
cylinder diagram with arbitrary boundary interactions.  

In general it will be quite involved to actually evaluate \factorb.  
However, it is trivial for a constant tachyon background because
then $Z_{\rm disk}(T)=e^{-T}Z_{\rm disk}(0)$ and so the sum over $I$ 
in \factorb\ can be evaluated using the on-shell result \factor .
It is appropriate to use the disk amplitude with the subtraction
of the infinite M\"{o}bius volume, {\it i.e.} the tree level action 
\bthree . 
This disk amplitude is just the tree level tachyon potential. 
We therefore find that the tachyon potential is loop 
corrected by
\eqn\applause{
V_{\rm loop} = \left(V(T)\right)^2{\tilde Z}_{\rm cyl}~.}
where ${\tilde Z}_{\rm cyl}$ was defined in \o\ and the tree
level potential is $V(T)=(1+T)e^{-T}$.

We can alternatively write our result \factorb\ as 
\eqn\alt{Z_{{\rm cyl}}(S_{{\rm bndy}}) = \int \! DX DY\,
\Psi(X)G(X,Y)\Psi(Y),}
where $\Psi(X) = e^{-S_{{\rm bndy}}(X)}$ represent the boundary state.
This defines $G(X,Y)$ as an off-shell closed string propagator.  It would
be interesting to compare with the version of the propagator studied 
in \cmnp.

\newsec{Comments}
In summary, we have discussed two proposals for the off-shell 
loop amplitude: the ``open string'' definition of section 3 
(if realizable) quantizes the tree level BSFT action directly 
and the ``closed string'' definition of section 4 demands
factorization in the closed string channel. We conclude
by comparing these definitions:

\item{(1)}
{\it The two approaches plausibly agree for on-shell external open string 
states.} Our closed string approach agrees with standard on-shell 
amplitudes by construction. Demonstrating this in the open string 
approach requires more analysis, but we see no reason in principle 
why this will not occur.

\item{(2)}
{\it They have different factorization properties}.  
By construction the open string approach will factorize correctly 
in the open string channel, and similarly for the closed string 
approach, but correct factorization in the opposite channel may 
not work off-shell. A similar situation pertains to cubic string 
field theory, where it was shown that off-shell open string
amplitudes have unphysical poles in the closed string channel \Freedman . 
It is not hard to understand where problems can arise. Closed string 
poles are high energy effects in open string field theory 
but when one goes off-shell and breaks conformal invariance on the 
world-sheet, renormalization of divergences can disturb this high 
energy behavior.

\item{(3)}
{\it They generate different amplitudes off-shell.} Consider amplitudes 
in the background of a constant tachyon. The open string action 
\expandb\ gives propagators weighted by $e^{T}$ and vertices weighted 
by $e^{-T}$ (times a possible linear term); so the loop expansion 
parameter in the open string approach is proportional to $e^{T}$ 
and a string diagram with $n$ boundaries picks up a factor 
$(e^T)^{n-1}$.  On the other hand, in the closed string approach 
the tachyon enters as $e^{-T}$ in each of the ``constituent'' disk 
amplitudes and the closed string propagators
are $T$ independent; therefore a diagram with $n$ boundaries is proportional
to $(e^{-T})^n$. These differences are most dramatic for large values 
of the tachyon, {\it i.e.} near the closed string vacuum. Here
the open string description is strongly coupled while the closed string 
description is weakly coupled.  

\item{(4)} 
Usually two prescriptions which agree on-shell but differ off-shell 
are related by a field redefinition. It is therefore tempting
to speculate that {\it an equivalence exists between the strongly coupled 
open string description and the weakly coupled closed string description}; 
in other words, there is a novel kind of duality. It would hardly be 
surprising that closed strings are the appropriate variables around the 
closed string vacuum. This type of relation would also be reminiscent 
of the AdS/CFT correspondence \refs{\MaldacenaRE,\GubserBC,\WittenQJ}. 
In any event, one lesson 
from AdS/CFT is that off-shell open strings can be dual to on-shell 
closed strings, and as such be observable. This may serve as further 
motivation for our insisting on factorization of the ``cylinder'' 
amplitude even with non-conformally invariant boundary interactions.
 
\bigskip\noindent
{\bf Note added:}
As this manuscript was being prepared, the preprint \refs{\BardakciCK} appeared
which contains the computation of sec 2.3.
 
\bigskip\noindent 
{\bf Acknowledgements:}
We thank R.~Gopakumar, E.~Martinec, L.~Rastelli, S.~Shatashvili, A.~Tseytlin, 
E.~Witten, B.~Zwiebach and particularly
D.~Kutasov for discussions. This work was supported in part by NSF
grant PHY-9901194 and by DOE grant
DE-FG0290ER-40560. FL was supported in part by a Robert R.
McCormick fellowship. FL thanks Rutgers University for hospitality
during parts of this work.

\listrefs

\end